\documentclass[pra,twocolumn,aps,superscriptaddress,longbibliography]{revtex4-1}

\usepackage{graphicx}
\usepackage{dcolumn}
\usepackage{bm}
\usepackage{epstopdf}
\usepackage{float}
\usepackage{amsmath}
\usepackage{xcolor}
\usepackage{comment}
\usepackage[normalem]{ulem}
\graphicspath{{./images/}}
\usepackage{pgfplots}

\usepackage{pgfplots}\pgfplotsset{compat=1.8}

\usepackage{tikz}
\usetikzlibrary{calc}

\usepackage{xcolor}
\definecolor{lightblue}{rgb}{0.2,0.2, 0.9}

\begin{document}
\title{Polaritons for  testing the universality of an impurity in a Bose-Einstein condensate}

\author{A. Camacho-Guardian}
\affiliation{Departamento de F\'isica Qu\'imica, Instituto de F\'isica, Universidad Nacional Aut\'onoma de M\'exico, Apartado Postal 20-364, Ciudad de M\'exico C.P. 01000, Mexico\looseness=-1}
\date{\today}
\begin{abstract}  Universality is a fundamental concept in physics that allows for the description of properties of systems that are independent of microscopic details. In this work, we demonstrate that the propagation of slow-light in the form of a dark-state polariton can encode universal aspects of an impurity strongly coupled to a Bose-Einstein condensate. This unveils a non-demolition scheme to probe impurity physics beyond the quasiparticle properties.  Based on a field theory that includes the two-body correlations at the exact level we demonstrate that under appropriate conditions, the damping rate of slow-light propagation reveals the high-energy universal tail in the polaron spectrum.
\end{abstract}
\maketitle
\maketitle

{\it Introduction.-} The concept of universality is ubiquitous in physics, referring to properties of systems that remain unchanged across interaction strengths, temperatures, and microscopic details of certain classes of phenomena. Ultracold gases have provided a platform in which to test universal features related to the properties of strongly correlated Fermi~\cite{Partridge2005, Schirotzek2008, Navon2010, Horikoshi2010, Stewart2010, Kuhnle2010,Bauer2014,Levinsen2017a} and Bose gases~\cite{Wild2012,Makotyn2014, Lopes2017, Eigen2017}.

The study of a quantum impurity coupled to a bosonic bath is a long-standing problem that dates back to Landau's and Pekar's studies on the behavior of electrons in a crystal~\cite{landau1948effective}. The understanding of the so-called Bose polaron has been renewed in light of new experiments with ultracold gases~\cite{hu2016bose, jorgensen2016observation, ardila2019analyzing, skou2021non, yan2020bose}, which has encouraged the development of new theoretical and numerical approaches~\cite{li2014variational, Christensen2015, Shchadilova2016, Levinsen2017, guenther2018bose, rath2013field, field2020fate, ardila2015impurity, grusdt2018strong, Drescher2020, Massignan2021, Guenther2021, Christianen2022, Christianen2022b, Yegovtsev2022} far beyond the regimes accessible in condensed matter systems. Despite this interest, many aspects of the Bose polaron remain elusive today. In this sense, the universal aspects of the Bose polaron are key features for understanding properties that are invariant to the microscopic details of the system. 

One of these properties is the universal frequency tail of the spectrum of the polaron, which is related to Tan's contact~\cite{tan2008energetics,tan2008large,tan2008generalized,Liu2020,Liu2020a} and provides direct information on the short-range interaction and initial dynamics of the polarons~\cite{Braaten2004, Fletcher2017, Yoshida2018, atoms9020022,skou2021non,Skou2022}. 

Polaritons, compound quasi-particles of light and matter, {have entered into the realm of strongly correlated phases of matter in condensed matter by serving as a powerful probe and measurement mechanism. Much of this sensing has exploited the ability of polaritons to interact via their matter component with a complex reservoir, forming new many-body states coined polaron-polaritons~\cite{efimkin2021electron, ravets2018polaron, cotlet2020rotons, imamoglu2021exciton, rana2020many, pimenov2017fermi, shahnazaryan2020tunable, Navadeh2019}. The recent progress in quantum gases has established a bridge that has been fruitful in transferring formalisms, numerical approaches, and knowledge between the fields of polarons in quantum gases and polaritons in semiconductor microcavities~\cite{Takemura2014,Sidler2017,cotlect2019transport,Levinsen2019,Bastarrachea2019,Camacho2021,Vashisht2022}.

 } In the context of atomic gases, dark-state polaritons have been studied in relation to polaron-polaritons, leading to the prediction of intriguing phenomena such as mass tuneable polaritons~\cite{Grusdt2016}, lossless polariton propagation above Landau's critical velocity~\cite{Nielsen2020}, and strongly interacting polaritons~\cite{Camacho2020}. Contrary to the progress made in condensed matter systems where polaritons are routinely utilized to investigate many-body phases of matter, this approach remains largely unexplored with ultracold gases. 

\begin{figure}[h]
\includegraphics[width=\columnwidth]{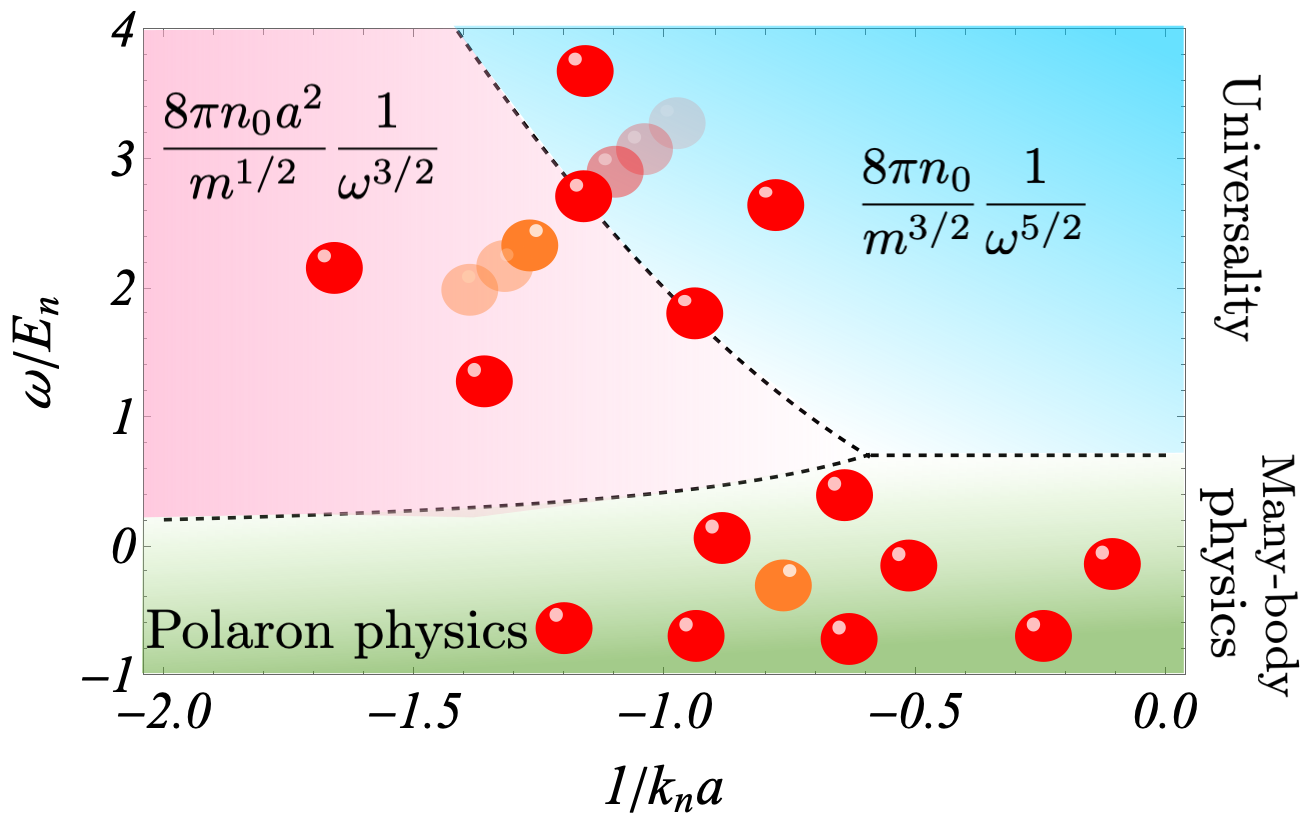}
\caption{Diagram of the spectral response of  an impurity coupled to a Bose-Einstein condensate as a function of the inverse of the impurity-boson scattering length and the frequency $\omega$. Universal high-energy tail spectrum  that scales with $\omega^{-5/2}$ (blue region) and weak-coupling universal regime $a^2\omega^{-3/2}$ pink region. Many-body correlations form within the green area.  Dashed line are guides to the crossover to the different scaling regimes  which are smooth transitions given by: blue to pink: $\omega=1/ma^2$, pink to green: $\omega=4\pi |a|n_0/m$ and green to blue: $\omega=(4\pi n_0/m^3)^{2/3}.$ {Here, the red and orange balls illustrate the majority and impurity atoms and the different energy regimes. }}
\label{Fig0}
\end{figure}

Polaritons offer alternative routes for addressing two critical aspects: a) Detecting strongly correlated states of quantum matter in a non-destructive manner, circumventing measurement schemes such as time-of-flight and b) studying polaron physics in the genuinely single-impurity limit by harnessing the capabilities of single-photon sources.

Motivated by this  open problem, in this article we propose polaritons in a BEC to measure the universal tail spectrum of the Bose polaron in ultracold gases. A field-theoretical approach, which incorporates two-body correlations of impurity-boson scattering, demonstrates that the damping rate of the polariton carries information about the universal tail highlighting the potential of polaritons to investigate features of strongly correlated systems in atomic gases beyond the quasiparticle picture.

{\it Universal tail of the polaron.-} Before we enter into the study of slow-light and polaritons to test the universal aspects of an impurity let us first briefly revise the problem of the Bose polaron and its universal aspects.

Consider a system formed by bosonic atoms 
\begin{gather}
\hat H_b=\sum_{\mathbf k}\epsilon_{\mathbf k}\hat b_{\mathbf k}^\dagger\hat b_{\mathbf k}+\frac{g_{bb}}{2V}\sum_{\mathbf k,\mathbf k',\mathbf q}\hat b^\dagger_{\mathbf k+\mathbf q} \hat b^\dagger_{\mathbf k'-\mathbf q}\hat b_{\mathbf k'}    \hat b_{\mathbf k},
\end{gather}
where $\hat b_{\mathbf k}^\dagger$ creates a boson with momentum $\mathbf k$ and energy $\epsilon_{\mathbf k}=k^2/2m$ where $m$ is the mass of the  bosons and $g_{bb}$ the boson-boson interaction strength. We assume that the bosons form a three-dimensional weakly Bose-Einstein condensate (BEC) correctly described by the Bogoliubov theory with a density $n_0.$  The energy and momentum scale associated to the BEC are $E_n=k_n^2/2m$ and $k_n=(6\pi^2n_0)^{1/3},$ respectively. We take $\hbar=1$.
 
 A single impurity interacts with the BEC, described by the term
 \begin{gather}
\hat H_{I}=\frac{g}{V}\sum_{\mathbf k,\mathbf k',\mathbf q}\hat b^\dagger_{\mathbf k+\mathbf q} \hat c^\dagger_{\mathbf k'-\mathbf q}\hat c_{\mathbf k'}    \hat b_{\mathbf k}, 
\end{gather}
where $\hat c_{\mathbf k}^\dagger$ creates an impurity atom with momentum $\mathbf k$. The strength of the interaction $g=4\pi a/m$ in terms of the boson-impurity scattering length that can be tuned on demand by means of a Feshbach resonance.  The volume of the system is denoted by $V$ whereas the kinetic energy of the c-impurities is described by the standard term $\hat H_c=\sum_{\mathbf k}\epsilon_{\mathbf k}^{(c)}\hat c^\dagger_{\mathbf k}\hat c_{\mathbf k}$  with $\epsilon_{\mathbf k}^{(c)}=k^2/2m+\epsilon_c$ being $\epsilon_c$ the energy off-set of the $c$-state with respect to the $b$-atoms.

The study of the many-body properties of the Bose polaron remains an outstanding problem that is far beyond the scope of this manuscript. Instead, in this article, we focus on the universal aspects of the Bose polaron, which appear regardless of the theoretical approach employed. {For an impurity immersed at  $t=0$ in a BEC, the spectral function is defined as $A(\mathbf 0,\omega)=\text{Re}\int_0^{\infty} \langle \psi_\mathbf 0|e^{-i\hat H t}|\psi_\mathbf 0\rangle e^{i\omega t}dt/\pi$, where $|\psi_\mathbf 0\rangle=\hat c^\dagger_{\mathbf 0}|\text{BEC}\rangle$. Here $\hat H=\hat H_b+\hat H_I+\hat H_c,$ and $|\text{BEC}\rangle$ the ground state of the bosons forming a BEC state, here the energy is referred to the non-interacting ground state. }

 We write the spectral function of the impurity as $A(0,\omega)=A_P(0,\omega)+A_{\text{tail}}(0,\omega)$, where $A_{\text{tail}}(0,\omega)$ represents the universal tail of the polaron spectrum and $A_P(0,\omega)$ relates to the many-body correlations that may form in the low-energy sector, which is not our current focus.  The tail of the spectrum is given by~\cite{Braaten2004} 
\begin{gather}
\label{tail}
A_{\text{tail}}(0,\omega)=\frac{8\pi n_0}{m^{3/2}\omega^{5/2}}\frac{m\omega a^2}{1+m\omega a^2}
\end{gather}
and holds for large $\omega.$ In Fig.~\eqref{Fig0}, we illustrate the different regimes of the polaron spectrum. It consists of a universal regime marked by the blue area that scales as $\omega^{-5/2}$. For weak interactions, a weakly interacting high-energy tail appears, marked by the pink area. The third regime, illustrated by the green area, corresponds to energies at which the polaron and many-body correlations form. Although there is still debate on the properties of the green region - that is, how the spectral function looks when many-body correlations arise- the tail of the polaron spectrum in Eq.~\eqref{tail} is an exact result that stems from the analytical expression for the two-body scattering
\begin{gather}
\label{tail0}
A_{\text{tail}}(0,\omega)= n_0 \frac{ A^{(0)}_{T}(\omega)}{\omega^2},
\end{gather}
{this equation, equivalent to Eq.~\ref{tail} } remarks that the tail of the polaron spectrum only relates to the two-body scattering properties. Where $A^{(0)}_T(\omega)=-2\text{Im}\mathcal T_v(0,\omega)$ is the spectral function of the  the two-body scattering matrix, in turn given by $\mathcal T_v(0,\omega)=\frac{4\pi a}{m}/(1+ia\sqrt{m\omega}).$ 

\begin{figure}[h]
\includegraphics[width=\columnwidth]{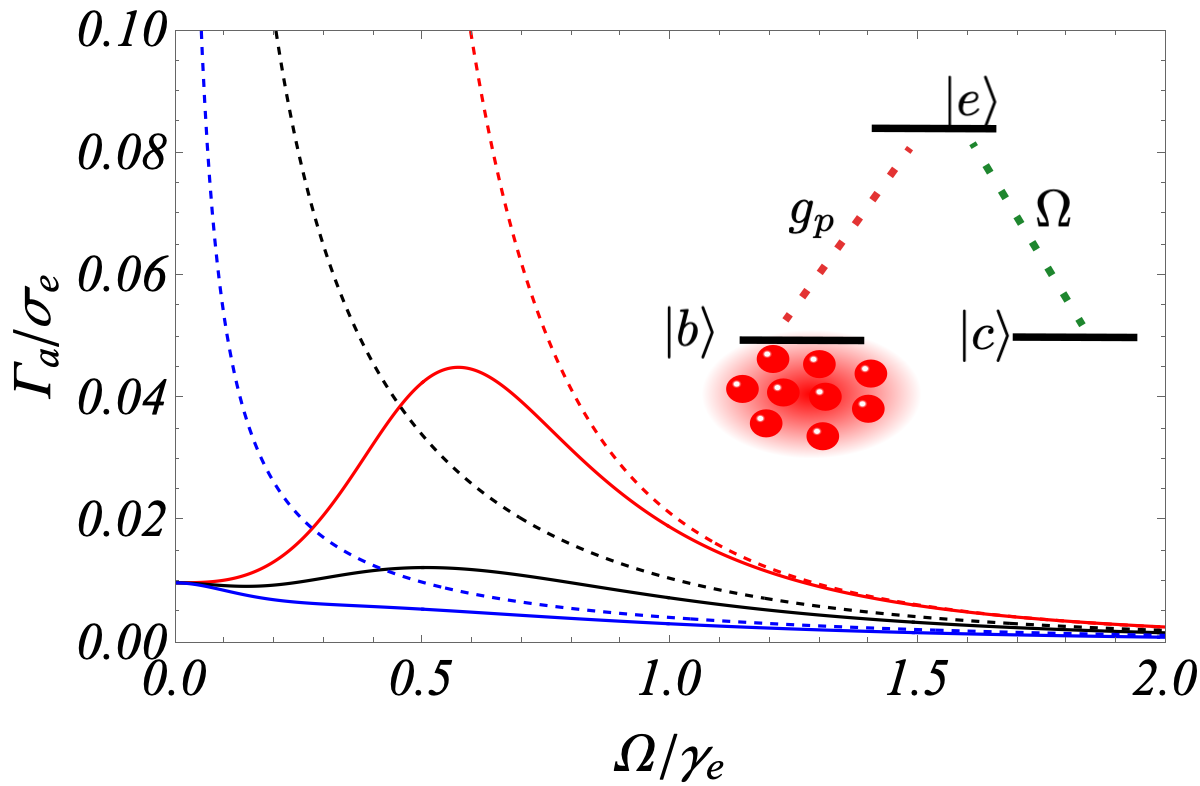}
\caption{The damping rate of the dark-state polaritons is plotted as a function of $\Omega/\gamma_e$ for several values of the inverse of the scattering length $1/k_na$. The red lines represent unitarity ($1/k_na=0$), with the solid curve giving the exact numerical result and the dashed line corresponding to Eq.~\eqref{univpolariton}. The black lines correspond to $1/k_na=-1$, with the solid line illustrating the full numerical solution and the dashed curve showing the asymptotic behavior. The blue lines represent weak interactions ($1/k_na=-2$). Inset: $\Lambda$ scheme of the atoms as detailed in the main text.  }
\label{Fig1}
\end{figure}

{\it Polaritons.-} We now turn our attention to the description of slow-light in the presence of atomic interactions. We consider atoms with an internal three-level structure as illustrated in Fig.~\eqref{Fig1} (inset) in a $\Lambda$ scheme. Atoms form a Bose-Einstein condensate in the $|b\rangle$ atomic state, an excited state $|e\rangle$ couples to a photon field $|a\rangle$ whereas a classical field couples the excited state $|e\rangle$ and a third state $|c\rangle.$ {In this case, the optical medium is the BEC, which is driven by a weak probe light denoted as $|a\rangle$. The absorption of a photon in the BEC excites an atom into the state $|e\rangle$. A stronger classical field then couples the excited state to the metastable state $|c\rangle$. As we will explain in the following, this configuration leads to EIT and dark-state polaritons.}

The coupling of the atoms in the BEC to the photons and excited state $|e\rangle$ is described by the standard light-matter Hamiltonian
 { \begin{gather}
\label{Hp0}
\hat H_{l-m}=\frac{g_p}{\sqrt{V}}\sum_{\mathbf k,\mathbf q}\left(\hat e^\dagger_{\mathbf k+\mathbf q}\hat a_{\mathbf k}\hat b_{\mathbf q}+\hat a^\dagger_{\mathbf k}\hat b^\dagger_{\mathbf q}\hat e_{\mathbf k+\mathbf q}\right)\\ \nonumber\approx\sqrt{n_0}g_p\sum_{\mathbf k}\left(\hat e^\dagger_{\mathbf k}\hat a_{\mathbf k}+\hat a^\dagger_{\mathbf k}\hat e_{\mathbf k}\right). 
\end{gather}
In the second line, we make use of the fact that for a BEC, the ground state is macroscopically populated, with $\hat b_{\mathbf 0} \approx \hat b^\dagger_{\mathbf 0} = \sqrt{N_0}$, where $N_0$ represents the number of condensate atoms.} The single photon coupling is denoted by $g_p.$ The ideal dispersion of the photons and the excited state are given by $\hat H_a=\sum_{\mathbf k}\omega_{\mathbf k}\hat a^\dagger_{\mathbf k}\hat a_{\mathbf k}$ and $\hat H_e=\sum_{\mathbf k}\epsilon_{\mathbf k}^{(e)}\hat e^\dagger_{\mathbf k}\hat e_{\mathbf k}$ respectively.  As usual, the photons have a linear dispersion $\omega_k=ck$ with $c$ the speed of light, while the dispersion of the excited state $\epsilon_{\mathbf k}^{(e)}=k^2/2m+\epsilon_e-i\gamma_e,$ where $\epsilon_e$ is the energy of the excited state relative to the $b$-state and $\gamma_e$ its damping rate. 
The $\Lambda$-scheme is completed by a classical control 
 field that couples the excited and metastable state and is given by the Hamiltonian
\begin{gather}
\hat H_{cl}=\Omega\sum_{\mathbf k}(\hat c^\dagger_{\mathbf k-\mathbf k_{\text{cl}}}\hat e_{\mathbf k}e^{i\omega_{\text{cl}t}}+\hat e^\dagger_{\mathbf k}\hat c_{\mathbf k-\mathbf k_{\text{cl}}}e^{-i\omega_{\text{cl}}t}  ), 
\end{gather}
with $\omega_{\text{cl}}$ and $\mathbf k_{\text{cl}}$the energy-momentum of a classical field and $\Omega$ the Rabi frequency.

To treat the light-matter coupling together with the atomic interactions we introduce the imaginary Green's function~\cite{Fetter1971} $$G_{\alpha\beta}(\mathbf k,\tau)=-\langle T_{\tau}\{\hat\psi_{\alpha\mathbf k}(\tau)\hat\psi^\dagger_{\beta\mathbf k}(0)\}\rangle,$$
where the indices $\alpha,\beta=\{c,e,a\}$ for the metastable, excited, and photon field respectively, here $\tau$ is the imaginary time and $T_{\tau}$ the time-ordering operator. {Our formalism allows us to explore finite-temperature and consider a finite density of impurities, in this work, however, we restrict our calculations to zero temperature and a single polariton in the BEC. }
The Dyson's equation in energy-momentum space governing the photon field is
\begin{gather}
\label{photonG}
G^{-1}_{aa}(\mathbf k,\omega)=G^{(0)}_{aa}(\mathbf k,\omega)^{-1}-    \Sigma_{aa}(\mathbf k,\omega)
\end{gather}
with the bare photonic Green's function $G^{(0)}_{aa}(\mathbf k,\omega)^{-1}=\omega-c|\mathbf k|$. 
 First, the diagonal self-energy of the photons is given by
\begin{gather}
\label{sigmagg}
\Sigma_{aa}(\mathbf k,\omega)=n_0g_p^2G_{ee}(\mathbf k,\omega),
\end{gather}
 The Green's function of the excited state $|e\rangle$ is 
\begin{gather}
\label{Gcc1}
G^{-1}_{ee}(\mathbf k,\omega)= G^{(0)}_{ee}(\mathbf k,\omega)^{-1}-\Omega^2G_{cc}(\mathbf k-\mathbf k_{cl},\omega),   
\end{gather}
where  $G^{(0)}_{ee}(\mathbf k,\omega)^{-1}=\omega-\epsilon_{\mathbf k}^{(e)},$ includes the induced damping and energy shifty due to the coupling to the $b$ state as described by the  Weisskopf-Wigner theory~\cite{weisskopf1997berechnung}. We consider a weak probe photon with momentum $\mathbf p_{r}$ we introduce the one- photon detuning $\Delta=c|\mathbf p_{r}|-\text{Re}\epsilon_{\mathbf p_r}^{(e)},$ and define the two-photon detuning as $\delta=\epsilon_c+\omega_{\text{cl}}-cp_r.$
 
In the absence of atomic interactions, we can use the {\it bare} propagator of the $c$-state in Eq.~\eqref{Gcc1}: $G^{-1}_{cc}(\mathbf k,\omega)=\omega-\epsilon_{\mathbf k}^{(c)}.$ In the rotating frame $\hat c^\dagger_{\mathbf k}\rightarrow \hat c^\dagger_{\mathbf k}e^{i\omega_{\text{cl}}t}$ , the energy of the $c$ state is given by $\epsilon_{\mathbf k}^{(c)}=k^2/2m+\delta+cp_r.$ In this case, when the two-photon detuning vanishes, light propagates in the form of a dark-state polariton~\cite{Fleischhauer2000, Fleischhauer2002,hau1999light}, which has a dramatically reduced group velocity of $v_g/c=1/(1+n_0g_p^2/\Omega^2)$ and propagates without losses in an opaque medium, a phenomenon also coined electromagnetically induced transparency (EIT)~\cite{boller1991observation,fleischhauer2005electromagnetically}. In our study, we only consider $\delta=0.$

We now include the boson-impurity interactions. Although the predicted state of the polaron may depend on the theoretical framework used, in this study, we focus only on the high-energy spectrum of the polaron. Therefore, we account for a formalism that includes this high-energy tail exactly using the so-called ladder approximation~\cite{rath2013field}. Within this approximation, the self-energy of the $c$-Green's function is given by
\begin{gather}
\Sigma_{cc}(\mathbf k,\omega)=n_0\mathcal T(\mathbf k,\omega). 
\end{gather}
Here, the scattering matrix $\mathcal T$ describes the interaction between a single  polariton  and a boson of the BEC, that is,
{
\begin{gather}
\label{Tpol}
\mathcal T(\mathbf k,\omega)=\frac{1}{\frac{m}{4\pi a}-\Pi(\mathbf k,\omega)},   
\end{gather}}
with 
\begin{gather}
\label{EcPair}
\Pi(\mathbf k,\omega)=-\sum_{\mathbf p,i\omega_{\nu}}G_{\text{BEC}}(\mathbf p,i\omega_{\nu})G^{(0)}_{cc}(\mathbf k-\mathbf p,\omega-i\omega_{\nu}). 
\end{gather}
{This approach is called the non-self consistent T-matrix (NSCT) approximation, it accounts for the infinite sum of diagrams in the ladder approximation: forward boson-impurity scattering~\cite{rath2013field}. This approximation takes into account that the impurity can only excite one Bogoliubov mode out of the condensate, and is equivalent to the bosonic version of the so-called Chevy's ansatz.
}   
Here $G_{\text{BEC}}^{-1}(\mathbf k,\omega)=\omega-k^2/2m$ corresponds to the Green's function of the BEC which for simplicity we assume to be ideal,  and $i\omega_\nu$ is a bosonic Matsubara frequency.  On the other hand, $G_{cc}(\mathbf k,\omega)$ corresponds to the $c$ propagator, which is now, dressed by the light-matter coupling 
\begin{gather}
\label{GcD}
[G^{(0)}_{cc}(\mathbf k-\mathbf k_{\text{cl}},\omega)]^{-1}=\omega-\epsilon^{(c)}_{\mathbf k-\mathbf k_{\text{cl}}}-\frac{\Omega^2}{\omega-\epsilon^{(e)}_{\mathbf k}-\frac{n_0g_p^2}{\omega-ck}}.
\end{gather}

The scattering matrix depends therefore on the coupling between the different states to light, and as we shall reveal, the properties of the $\mathcal T$-matrix reflect on the propagation of the dark-state polaritons.

{\it Probing polaron universality with polaritons.-} Before delving into technical and numerical details, let us first discuss some limits that will guide the physical interpretation of our results.

Under perfect EIT conditions, light can propagate without losses in an otherwise opaque BEC. In this study, we use the losses induced by atomic interactions on the dark-state polariton to probe the universal tail of the polaron spectrum. We begin by defining the damping rate of the polariton, which will serve as a starting point for our analysis
\begin{gather}
\Gamma_a=-Z_a\text{Im}\Sigma_{aa}(\mathbf p_r,cp_r),
\end{gather}
where $Z_a$ is the residue of the polariton $Z_a=1/(1+\frac{n_0g_p^2}{\Omega^2}).$ 

The main result of our work is summarised in the following equation
\begin{gather}
\label{univpolariton}
\Gamma_a\approx A_{\text{tail}}\left(0,\omega_l\right)\omega_l^2,
\end{gather}
where $\omega_l=-\Omega^2/\Delta$ with $\Delta<0$. This expression relates the losses of the polariton to the universal tail of the polaron spectrum. It provides a non-demolition probing scheme, allowing for the measurement of the high-energy spectrum of a strongly interacting impurity coupled to a BEC via polaritons. That is, by tuning the light-matter parameters $\Omega$ and $\Delta$, one can measure the tail of the polaron spectrum as in Eq.~\eqref{univpolariton}. Equation~\eqref{univpolariton}  links the tail of the two-body scattering to the damping rate of the polariton as it can be re-written as $\Gamma_a\approx n_0A^{(0)}_{T}\left(0,\omega_l\right).$ 

The physical origin and validity of our main result in Eq.~\eqref{univpolariton} rely on two considerations.

First, from Eq.~\eqref{photonG} evaluated for an incoming photon with energy-momentum $(\mathbf p_r,cp_r)$, we have
\begin{gather}
G^{-1}_{aa}(\mathbf p_r,cp_r)=-\frac{n_0g_p^2}{\Delta+i\gamma_e+\frac{\Omega^2}{n_0\mathcal T(\mathbf 0,cp_r)}}.
\end{gather}
We observe that if the polaron energies which are of the order of $n_0\mathcal T(0,cp_r)$ remain smaller than the width of the EIT, i.e., $\frac{\Omega^2}{\sqrt{\gamma_e^2+\Delta^2} n_0\mathcal T(\mathbf 0,cp_r)}\gg 1$, then we can estimate the imaginary part of the dressed photon propagator as $\text{Im}G^{-1}_{aa}(\mathbf p_r,cp_r)\approx \frac{n_0g_p^2}{\Omega^2} n_0\text{Im}\mathcal T(0,cp_r)$.

Second, when the one-photon detuning is also larger than the typical energies of the impurity, the Green's function of the $c$ atoms can be approximated as $[G^{(0)}_{cc}(\mathbf k-\mathbf k_{\text{cl}},\omega)]^{-1}\approx \omega-\epsilon^{(c)}_{\mathbf k-\mathbf k_{\text{cl}}}-\frac{\Omega^2}{\Delta}$ in Eq.\eqref{GcD}. This approximation allows us to replace the polariton-boson scattering matrix in Eq.\eqref{Tpol} with the impurity-boson scattering matrix evaluated at energy $\omega_l$, i.e., $\mathcal T(\mathbf 0,cp_r)\rightarrow \mathcal T_\nu(\mathbf 0,\omega_l)$. Under these two conditions, we obtain Eq.~\eqref{univpolariton}. We emphasize that in our numerical calculations, we retain the full Green's function for the impurity and photon.


To test numerically our theoretical analysis, we now obtain the photon propagator numerically.   For relevant EIT experiments~\cite{lampis2016coherent}, the energy of the $c$-state is typically on the order of $\epsilon_c\sim 100-400\text{ MHz}$ above the energy of the $b$-state. Achieving a vanishing two-photon detuning requires $\epsilon_c+ck_{\text{cl}}-cp_r=0$. By rewriting $k_{\text{cl}}=p_r+\Delta p$, we find that $\Delta p/k_n\sim 10^{-6}$ for current polaron experiments ($k_n\sim 1/1000a_0$, with $a_0$ being the Bohr radius)~\cite{hu2016bose}. In other words, due to the steep dispersion of light, the momentum shift needed to fix $\delta=0$ for a realistic value of $\epsilon_c$ is completely negligible, $\Delta p/k_n\ll 1$. Consequently, we can effectively probe the zero-momentum polaron. In addition, we take  $\gamma_e=10$MHz, a one-photon detuning of $\Delta=-10\gamma_e$,  $k_{\text{cl}}=0.1k_n,$ a BEC with $E_n=500\text{kHz},$  and $\sqrt{n_0}g_p/\gamma_e=30,500$.

 {To understand our results, let us link the damping rate $\Gamma_a$ to a physical observable}, we note that a photon propagating in a BEC with $\Omega=0$ spends an amount $t^{(0)}_a=L/c$ of time in the condensate of length $L$. Thus, it is convenient to introduce the dimensionless quantity termed optical depth $\text{OD}_0=\Gamma^{(0)}_a t^{(0)}$  where $\Gamma_a^{(0)}=n_0g_p^2/\gamma_e,$  is the damping rate of the photon in the two-level medium ($\Omega=0).$ Comparing the same quantity for slow-light $\text{OD}=\Gamma_at_a$ and $t_a=L/v_g$ we obtain that the ratio  $\text{OD}/\text{OD}_0=\Gamma_a/\sigma_e$ can be written in terms of $\sigma_e=\Omega^2/\gamma_e,$ which is the normalization we will use. Under strict EIT conditions $\Gamma_a=0$.

Figure~\eqref{Fig1} shows the damping rate of the dark-state polariton as a function of $\Omega$ for different values of the boson-impurity scattering length. The solid lines are the result of the full numerical solutions, while the dashed line corresponds to the asymptotic tail in Eq.\eqref{univpolariton}. Overall, we see that  the polariton damping agrees well with the high-energy tail of Eq.\eqref{univpolariton}.
 \begin{figure}[h]
\includegraphics[width=\columnwidth]{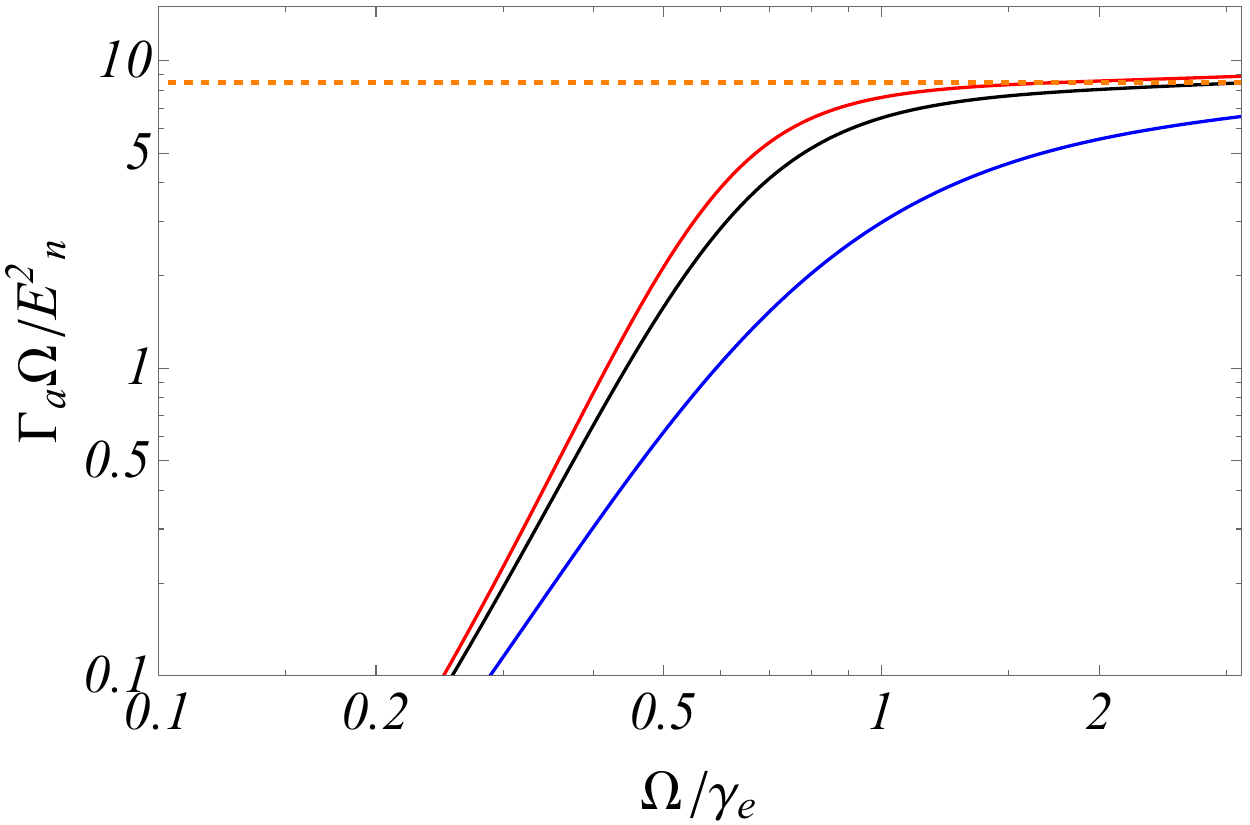}
\caption{ Universail tail from the damping rate of polaritons. In red we show the unitary regime $1/k_na=0$, in black we have $1/k_na=-0.2$ while the blue line gives $1/k_na=-1$. Solid curves give the numerical results, the orange line gives the universal limit. The remaining parameters as in Fig.~\eqref{Fig1}.}
\label{Fig2}
\end{figure}
For strong interactions, in particular, at unitarity $1/k_na$ (red lines), we observe that the polariton damping joins the universal tail at $\Omega/\gamma_e\approx 1$ and seems to agree very well with the tail for large $\Omega$. On the other hand, as $\Omega$ is decreased, we obtain significant deviations from Eq.~\eqref{univpolariton} until the comparison no longer makes sense. If at unitarity the typical polaron energies are of the order of $E_n$, for $\Omega/\gamma_e=1.0$, we obtain that for the parameters taken, $\Omega^2/(\sqrt{\Delta^2+\gamma_e^2}E_n)=2$. Thus, in accordance with our analysis, we expect to start agreeing with the universal tail close to this value and higher values of $\Omega$.

At intermediate coupling strengths ($1/k_na=-1$), we also observe agreement between the high-energy tail and the numerical results. However, the agreement shifts to larger values of $\Omega$. Physically, this can be understood from Fig.~\eqref{Fig0}: as the interaction strength decreases, the universal tail shifts to even higher energies. Thus, in our case, we are required to tune $\Omega$ to higher values.

Finally, for weak interactions ($1/k_na=-2$), we find that as the universal tail is pushed further in energies, the agreement between the numerics and Eq.~\eqref{univpolariton} occurs only at large values of $\Omega$, consistent with our previous discussion.

Our analysis indicates that the damping rate of the dark-state polariton scales with the inverse of the Rabi frequency, $\Gamma_a\approx n_0/\Omega$. However, it is difficult to directly confirm this hypothesis from Fig.~\eqref{Fig1}. In order to demonstrate this relationship, we present $\Gamma_a\Omega$ on a logarithmic scale in Fig.~\eqref{Fig2}. We observe that for strong interactions, the value of $\Gamma_a\Omega$ approaches the unitary value of $\Gamma_a\Omega\approx\frac{n_0}{m^{3/2}}\sqrt{|\Delta|}$ (dashed orange line). At unitarity (solid red line), we observe an evident change of slope close to $\Omega/\gamma_e=1$, when $\Omega^2/(\sqrt{\Delta^2+\gamma_e^2}E_n)=1$. As the interaction is decreased, we find that the high-energy tail is reached at larger values of $\Omega$, as discussed previously.

Our result unveil an intriguing and non-trivial link between the universal aspects of the polaron and slow-light. It demonstrates that the damping rate of the polariton encodes the high-energy universal tail of the impurity,  and can be used as  as a non-demolition probing scheme, {that is, procedures like time-of-flight where the BEC is {\it destroyed} during the measurement protocol are avoided with polaritons}. Unlike polaron experiments with highly-imbalanced population mixtures, where the signal of the impurity typically scales linearly with the density of the impurities,  {state-of-the-art experiments with polariton in quantum gases with single photons~\cite{Liang2018}, suggest that our proposal can test genuinely single-impurity physics.   }

{\it Conclusions.-}  
In this article, we have shown that the damping rate of the polariton can be used to probe the high-energy energy spectrum of a quantum impurity coupled to a BEC. Our results relate the propagation of slow-light to the universal properties of an impurity strongly interacting with its surroundings beyond the quasiparticle picture. This work suggest polaritons as a non-demolition probing scheme that can test single-impurity physics.

Our study reflects a rich interplay between well-established polariton physics and high-energy polaron physics.  The vast playground of these two phenomena may serve as a benchmark for future experiments and theories to explore polariton and polaron physics beyond the linear regime~\cite{Camacho2018a,Camacho2018b,Will2021,Camacho2022c}.

{\it Acknowledgments.-} A. C. G. thanks valuable discussions with G. M. Bruun which motivated this work and for valuable  comments to the manuscript.
A.C.G. acknowledges financial support from UNAM DGAPA PAPIIT Grants No. IA101923,  UNAM DGAPA PAPIME Grants No. PE101223 and PIIF23. 

\bibliography{references}
\end{document}